\documentclass{JHEP}
\usepackage{epsfig}
\usepackage{amssymb}

\newcommand{\bea}{\begin{eqnarray}}
\newcommand{\eea}{\end{eqnarray}}

\def\d{{\rm d}}

\preprint{SNUST 050701\\
{\tt hep-th/0507082}}
\title{Black Hole as Emergent Holographic Geometry \\
of Weakly Interacting Hot Yang-Mills Gas~\footnote{This work was
supported in part by the the MOST-KOSEF SRC Program CQUeST
(R11-2005-021).} }
\author{Soo-Jong Rey $^a$, Yasuaki Hikida $^b$\\
~~~~~~~~~~~~~~\\
${}^a$ School of Physics and Astronomy \& BK-21 Physics Division \\
Seoul National University, Seoul 151-747 {\rm KOREA}\\
$^b$ Theory Group, KEK, Tsukuba, Ibaraki 305-0801 {\rm JAPAN} \\
~~~~~~~~~~~~~~~~~~~\\
 \email{sjrey@snu.ac.kr, hikida@post.kek.jp} }
\abstract{We demonstrate five-dimensional anti-de Sitter black hole
emerges as dual geometry holographic to weakly interacting ${\cal
N}=4$ superconformal Yang-Mills theory. We first note that an ideal
probe of the dual geometry is the Yang-Mills instanton, probing
point by point in spacetime. We then study instanton moduli space at
finite temperature by adopting Hitchin's proposal that geometry of
the moduli space is definable by Fisher-Rao "information geometry".
In Yang-Mills theory, the information metric is measured by a novel
class of gauge-invariant, nonlocal operators in the instanton
sector. We show that the moduli space metric exhibits (1)
asymptotically anti-de Sitter, (2) horizon at radial distance set by
the Yang-Mills temperature, and (3) after Wick rotation of the
moduli space to the Lorentzian signature, a singularity at the
origin. We argue that the dual geometry emerges even for rank of
gauge groups of order unity and for weak `t Hooft coupling.}
\keywords{AdS/CFT correspondence, instanton, holography, black hole}
\begin{document}

\section{Background}

Maldacena's gauge-gravity correspondence \cite{maldacena} refers to
the remarkable relation between 4-dimensional conformal gauge theory
and 5-dimensional string theory on anti-de Sitter space. The gauge
theory is characterized by gauge group SU(N) and `t Hooft coupling
$\lambda^2 = g^2_{\rm YM} N$, while string theory is characterized
by coupling $g_{\rm st}$ and curvature scale (in string scale unit)
$R$ of the 5-dimensional anti-de Sitter space, whose metric in
Poincar\'e coordinates is given by
\bea \d s^2_0 = R^2 \left({1 \over u^2} \left[ - \d t^2 + \d {\bf
x}^2 \right] + {\d u^2 \over u^2}\right). \label{1} \eea
The correspondence is valid in the limit the string theory is weakly
coupled and anti-de Sitter space is nearly flat, and identifies
coupling parameters on each side as
\bea  N \sim {1 \over g_{\rm st}} \gg 1 \qquad \mbox{and} \qquad
\lambda^2 \sim R^4 \gg 1. \label{2} \eea
The correspondence is a new kind of strong-weak coupling duality,
and has been established firmly in the limit of (\ref{2}).

Maldacena's correspondence is extendible to Yang-Mills theory at
finite temperature. The dual string theory is then defined on
Schwarzschild black hole in 5-dimensional anti-de Sitter space:
\bea \d s^2_T = R^2  \left[ - {1 \over u^2} \left( 1 - {u^4 \over
u_o^4} \right) \d t^2 + \left( 1 - { u^4 \over u_o^4}\right)^{-1}
{\d u^2\over u^2}+ {1 \over u^2} \d {\bf x}^2 \right], \label{3}\eea
whose surface gravity at the horizon $u = u_0$ set by the Yang-Mills
temperature $T$. The Maldacena's correspondence, as is formulated,
is valid for large $N$ and strong `t Hooft coupling, see (2). On the
other hand, enormous insight on gauge and string theories would be
gained by understanding other regime of the coupling parameters. So,
we consider taking extreme opposite regime
\bea N \sim {\cal O}(1) \qquad {\rm and} \qquad \lambda \ll 1,
\label{4}\eea
and pose the following questions. Despite being in the regime of
strongly interacting string theory, can a black hole geometry (3) or
close to it be reconstructed out of weakly interacting Yang-Mills
gas? Can we explore interior of the horizon and see if the black
hole singularity resolved by strong stringy and quantum effects? If
resolved, what are universal features of the resolved geometry?
Undoubtedly, affirmative answers to these questions would yield far
reaching consequences to our ultimate understanding on gauge theory,
quantum gravity, string theory, black hole, and (spacelike)
singularity.

These questions are interesting, but one may subscribe a certain
doubt because Eq.(\ref{4}) lies outside the validity range of
Maldacena's correspondence. On the other hand, there are physical
quantities whose behavior seems to support smooth interpolation
between weak and strong coupling regimes. Take the thermodynamic
free energy density ${\cal F}(N, \lambda)$ of hot Yang-Mills gas. At
large $N$ but at arbitrary $\lambda$, the free energy density
behaves as ${\cal F}(N, T) = h(\lambda) \left( -{1 \over 6} \pi^2
N^2 T^4 \right)$. At strong `t Hooft coupling regime, $\lambda \gg
1$, the leading-order correction was computed from string worldsheet
corrections to the Type IIB supergravity and hence to the black hole
geometry. It shows that $h(\lambda)$ increases monotonically from
3/4 as $\lambda$ is decreased \cite{klebanov}. At weak `t Hooft
coupling regime, $\lambda \ll 1$, the leading-order corrections was
computed up to two loop by thermal perturbation theory for
Yang-Mills theory, taking into account of bubble resummation and
electric screening thereof. With Pad\'e approximation, it was shown
that $h(\lambda)$ decreases monotonically from 1 as $\lambda$ is
increased \cite{kimrey}.

In this work, we shall study ${\cal N}=4$ super Yang-Mills theory at
finite temperature at weak coupling and look for a signature, if
any, of holography \cite{thooft} and of emergent black hole
geometry. A moment of thought indicates that perturbative gluon
dynamics at weak coupling is an unlikely setup to discover such
geometry, and, by S-duality, semiclassical solitons (such as
instantons or monopoles) at strong coupling limit is likewise an
unlikely setup. We will therefore focus on instantons in weak
coupling limit, where semiclassical treatment is well justified.
Morally speaking, this is S-dual to gluon dynamics at strong
coupling limit --- the situation where Maldacena's correspondence
was extensively confirmed. Moreover, instanton in Yang-Mills theory
is the counterpart of D-instanton in Type IIB string theory on 5d
anti-de Sitter space. The D-instanton is an ideal probe since it has
sub-string scale size at weak coupling and it probes the spacetime
point by point. Built upon a prescription by Hitchin \cite{hitchin},
we will show that a kind of black hole geometry in asymptotic
anti-de Sitter background emerges as the "information geometry"
\cite{info1, info2}  for the moduli space of Yang-Mills instanton at
finite temperature.

\section{Theoretical procedure}

The Yang-Mills instanton for gauge group $G=$SU(2) is a self-dual
configuration of the field strength $F^a_{mn}$: $F^a_{mn} = {1 \over
2} {\epsilon_{mn}}^{pq} F^a_{pq}$. For a single instanton, the
Yang-Mills field strength and Lagrangian is given by
\bea F^a_{mn} = - \eta_{mn}^a {\rho^2 \over [(x-a)^2 + \rho^2]^2},
\qquad \mbox{and} \qquad  {\cal L}_{\rm YM} &\equiv& {1 \over 2}
{\rm Tr} F^2_{mn} = {\rho^4 \over [(x - a)^2 + \rho^2]^4}. \label{5}
\eea
Notice that ${\cal L}_{YM}$ is a function of both four-dimensional
coordinates $x^m$ on $\bf{R}^4$ and five-dimensional coordinates
$Z^A \equiv (a^m, \rho)$ (which parametrize center and size of the
instanton) on the instanton moduli space ${\cal M}$ whose topology
is $\bf{R}_4 \times \bf{R}_+$.

The geometry of ${\cal M}$ is definable by an appropriate choice of
the metric. A familiar choice, studied extensively, is the
`$L^2$-metric' $G^{L^2}_{AB} := \int_{\mathbb{R}_4} g^{mn}(x) \,
(\partial_A A_m \partial_B A_n ) (x; Z)$. It yields, for the base
manifold $\bf{R}^4$, $\d s^2 (G^{L^2}) = \d \rho^2 + \d a^2 = \d Z^2
$.
As pointed out in \cite{hitchin}, the $L^2$-metric is not convenient
for studying {\sl differential geometry} of the moduli space ${\cal
M}$, since the metric does not display the underlying conformal
invariance and the moduli space is geometrically incomplete as the
small instanton singularity $\rho = 0$ is at a finite distance from
a generic point on ${\cal M}$. Also, the metric depends sensitively
on topology and choice of the metric on the base manifold $B$, and
for $G =$SU(N), one instanton moduli space is 4N-dimensional,
depending on the rigid SU(N) gauge transformations. In the context
of Maldacena's gauge-gravity correspondence, it is worthwhile to
recall that the ADHM instanton for ${\cal N}=4$ super Yang-Mills
theory \cite{dorey} yielded $AdS_5 \times \mathbb{S}_5$ {\sl after}
integrating out the zero-modes for U(N) global gauge rotations by
the large $N$ saddle point approximation. As a result, the
computation was reduced entirely to $G=$SU(2) instanton calculus.

To keep the underlying symmetry manifest and better aid differential
geometric properties, Hitchin \cite{hitchin} proposed an alternative
definition of the moduli space geometry. For the set of instantons
with a fixed instanton number $Q$, the Lagrangian density ${\cal
L}_{\rm YM}$ may be considered as a probability distribution
functions: $\int_{\mathbb{R}^4} {\cal L}_{\rm YM} = Q$.
Hitchin then proposes to utilize so-called Fisher-Rao's information
metric to describe the geometry of the instanton moduli space ${\cal
M}$. Extending his prescription, we propose the information metric
in gauge theory in terms of quantum average of nonlocal, gauge
invariant operator:
\bea G_{AB}^{\rm info} (Z) \equiv \int_{\mathbb{R}^4} \langle {\cal
L}_{\rm YM} \, (\partial_A \log {\cal L}_{\rm YM}) (\partial_B \log
{\cal L}_{\rm YM}) \rangle. \label{metric} \eea
Here, the bracket abbreviates normalized functional integral over
the gauge field configurations (whose definition involves the
standard $L^2$-metric for the functional integral measure). At
leading order in weak coupling regime, one simply evaluates
(\ref{metric}) for the instanton solution, obtaining Hitchin's
original prescription.

The information metric (\ref{metric}) probes variation of the
instanton density over the moduli space ${\cal M}$. Indeed, from
(\ref{5}), information metric of the unit charge, SU(2) Yang-Mills
instanton is readily extracted as
\bea \d s^2 (G^{\rm info}) = c { \d \rho^2 + \d a^2 \over \rho^2}.
\label{0t} \eea
So, identifying $\rho, a^m$ with $u, x^m$ in (2), we constructed
5-dimensional Euclidean anti-de Sitter space ${\bf H}^5$ as the
emergent geometry of weakly coupled Yang-Mills theory. Notice that
the information metric (\ref{0t}) exhibits underlying conformal
invariance manifestly, geodesic completeness (small instanton
singularity is at infinite distance from a generic point in the
interior), and invariance under the global gauge rotation.

The above result may be interpreted as follows. In Maldacena's
gauge-gravity correspondence, classical equation of motion of
dilaton $\Phi$, as derived by the variation of the supergravity
effective action $W_{\rm sugra}$, is sourced by the Lagrangian
density of the ${\cal N}=4$ super Yang-Mills theory \cite{corr1,
corr2}. Thus, for an instanton configuration of a fixed topological
charge,
\bea \left( {\delta W_{\rm sugra} \over \delta \Phi} \right)_{\rm
D-instanton} = {1 \over g^2_{\rm YM}} \langle {\cal L} \rangle_{\rm
instanton}. \eea
As seen above, the right-hand side depends not only on the
coordinates of the base manifold $B=\bf{R}^4$ but also on the
coordinates on ${\cal M}$. Therefore, the Lagrangian density for the
instanton, equivalently, topological charge density is interpretable
as the bulk-to-boundary propagator for a massless spin-0 field. In
fact, one can show that
\bea \Box_{\mathbb{H}_5} {\cal L}_{\rm YM}(x, Z) = 0 \qquad
\mbox{and} \qquad  {\rm lim}_{\rho \rightarrow 0} {\cal L}_{\rm
YM}(x, Z) = \delta^{(4)}(x - x_0), \nonumber \eea
both of which are precisely the defining equations for the
bulk-boundary propagators in AdS space. One can further construct
bulk-boundary propagators for {\sl massive} spin-0 fields by raising
power of the instanton charge density ${\cal L}_{\rm YM}$:
\bea \Box_{\mathbb{H}_5} \Big( {\cal L}_{\rm YM} (x, Z)\Big)^{m^2/4}
= m^2 (m^2 -4) \Big( {\cal L}_{\rm YM}(x, Z) \Big)^{m^2/4}. \eea
The rate of variation of the instanton density over the moduli space
${\cal M}$ is measurable by the logarithmic derivative $v_A (x;Z)
\equiv
\partial_A \log {\cal L}_{\rm YM}(x:Z)$. By explicit computation
\cite{narain}, one can show that these logarithmic vector fields obey the
`Hamilton-Jacobi' equation $||v ||^2_{\rm info} \equiv G^{AB}_{\rm
info} v_A v_B = 1$
for {\sl every} point $x^m$ on the base manifold $B = \mathbb{R}^4$.
Therefore, $v_A$ may be interpreted as the velocity field, and
$G_{AB}^{\rm info}$ a viable metric defining the geodesic flows on
the moduli space ${\cal M}$.

Salient features of the information metric are \cite{narain}
\hfill\break
$\bullet$ Information metric $G_{\rm AB}$ is Einstein. Furthermore,
perturbation of information metric $G_{\rm AB}$ is the bulk-boundary
propagator for `graviton'. \hfill\break
$\bullet$ The instanton charge density ${\cal L}_{\rm YM}$ is the
bulk-boundary propagator for massless scalar fields. This relation
holds also for perturbation of ${\cal L}_{\rm YM}$. \hfill\break
$\bullet$ The geodesic distance, once properly regularized, between
the boundary point $x^m$ of the base manifold $B = \mathbb{R}^4$ and
the bulk point $Z^A$ is given by log ${\cal L}(x:Z)$.

We shall now study the information geometry of instanton moduli
space for Yang-Mills theory at finite temperature $T$. In the
Matsubara formulation, the Euclidean time direction is topologically
$\mathbb{S}^1$ with periodicity $\beta = 2 \pi / T$. In general, the
self-dual Yang-Mills potential $A_\mu({\bf x}, t)$ on $\mathbb{R}^3
\times \mathbb{S}^1$ obeys the quasi-periodicity condition: $A_\mu
({\bf x}, t + \beta) = \omega^{-1} A_\mu ({\bf x}, t) \omega$ where
$\omega$ is an element of the SU(2) gauge group. By making the large
gauge transformation $A_\mu ({\bf x}, t) \rightarrow A_\mu^\Omega
({\bf x}, t) = A_\mu({\bf x}, t) + i \Omega ({\bf x}, t)
\partial_\mu \Omega^{-1}({\bf x}, t), \Omega = \omega^{t /
\beta}$,
we can bring the gauge potential strictly periodic. Semiclassically,
only those gauge field configurations with the {\sl same} gauge
orientation at infinity contribute to the functional integral. Such
periodic instantons, also known as calorons \cite{caloron}, can be
found from multi-instanton solution of `t Hooft type, in which gauge
orientation of constituent instantons is identical. We arrange the
constituent instantons on the same spatial location, same size, but
arrayed along the temporal direction with separation $\beta$.
Summing over the infinite constituent instantons, the caloron
solution is given by $A^a_m = - \eta^a_{mn} \partial_n \log
\Pi(x:Z)$ where the prepotential $\Pi$ is given by
\bea \Pi(x: Z) = 1 + {\rho^2 T^2 \over r T} {\sinh r T \over 2 (
\cosh  r T - \cos t T)} \eea
on base manifold $B = \mathbb{R}_3 \times \mathbb{S}_1$. For later
convenience, we also abbreviated $x=({\bf x}, t)$ and $Z = (\rho,
{\bf a}, a_o)$ and denoted $r \equiv |{\bf x} - {\bf a}|$, $t = (t -
a_0)$. Notice that the instanton size moduli $\rho$ refers to that
of the constituents, viz. of the zero temperature instantons.
The temperature $T$ is related to the Euclidean time periodicity
$\beta$ as $\beta = 2 \pi/T$. The prepotential $\Pi$ is manifestly
nonsingular, since the denominator never vanishes, except the
measure-zero configuration at the instanton center. In this case,
even though the prepotential diverges, integration over ${\bf R}^4$
renders it finite.

Anatomy of the caloron and its physical implications were studied
thoroughly \cite{grossetal}. At large distance, $|{\bf x}| \gg
\beta$, the prepotential is expandable as $\Pi(x:Z) \sim 1 + {\rho^2
T / 2 r} + {\cal O}(e^{-rT/2 \pi})$, so the temporal and the spatial
components of the gauge potential become
\bea A_0^a \sim -{x^a \over r^2} \left(1 + {2 r \over \rho^2 T}
\right)^{-1} + \cdots \qquad \mbox{and} \qquad A_i^a \sim
\epsilon^a_{ij} {x^j \over r^2} \left( 1 + {2 r \over \rho^2 T}
\right)^{-1} + \cdots. \eea
It then follows that the color electric and magnetic fields decay as
$r^{-1}$, and hence exhibit characteristics of magnetic monopoles.
Indeed, magnetic monopole is interpretable as an infinite array of
Yang-Mills instantons along the Euclidean time direction.

At short distance, $|{\bf x}| \ll \beta$, the prepotential is
expandable as $\Pi (x: Z) \sim ( 1 + \rho^2 T^2 /12 ) + {\rho^2 /
|x|^2} + \rho^2 {\cal O}(|x|^2/\beta)$. Consequently, the gauge
potential asymptotes to
\bea A_m^a \sim \eta^a_{mn} \, {\rho^2_{T} \over |x|^2} \, {x^n
\over (|x|^2 + \rho^2_{T})}+ \cdots \eea
where $\rho_{T}$ is related to $\rho$ as
\bea {1 \over \rho^2_{T}} = {1 \over \rho^2} + {T^2 \over 12}.
\label{caloronrelation} \eea
Notice that the gauge potential takes precisely the same form as the
zero temperature Yang-Mills instanton except the crucial difference
that the size moduli ought to be identified with $\rho_T$, not
$\rho$ itself. It is important to recall that the asymptotic
expansion is valid for {\sl any} value of $\rho$. It then follows
that, as interpreted in terms of zero temperature Yang-Mills
instanton, the caloron size moduli is limited over the range:
\bea 0 \le \rho \le \infty \qquad \longrightarrow \qquad 0 \le
\rho_{T} \le {2 \sqrt{3} \over T}. \label{range}\eea
The fact that the asymptotic caloron behaves as the zero temperature
instanton implies that the caloron ought to exhibit conformal
symmetry asymptotically near $\rho_T \sim 0$. Beyond the leading
semiclassical approximation, generically there will be corrections
arising from logarithmically running coupling constants. For
conformally invariant ${\cal N}=4$ Yang-Mills theory, such effects
are absent, rendering the conformal symmetry at $\rho_T \sim 0$
exact.

The result, Eq.(\ref{range}) is elementary but leads to remarkable
consequences. We expect that, as measured in scales of the zero
temperature instanton, the caloron size does not grow forever (which
was so at zero temperature) as $\rho \rightarrow \infty$. Rather,
caloron size saturates at a finite size set by the temperature $T$.
It suggests us to interpret the maximum caloron size as the emergent
horizon of the black hole in the background of the information
geometry space ${\bf H}_5$ in (\ref{0t}).

\section{Analysis and Results}

Having understood the caloron configuration, we would like to
understand the geometry of caloron's moduli space. As for the zero
temperature instantons, we shall adopt Fisher-Rao's information
metric (\ref{metric}) as the definition of the geometry. In terms of
the prepotential $\Pi(x:Z)$, the caloron action density is
expressible by ${\cal L}_{\rm YM} = {1 \over 2} \Box_{\mathbb{R}_4}
\left(\partial_m \log \Pi \right)^2 = - \Box^2_{\mathbb{R}_4} \log
\Pi$. We anticipate the information metric for the caloron of fixed
topological charge takes the form
\bea \d s^2_{\rm caloron} = G_{tt}[u] \d t^2 + G_{uu}[u] \d u^2 +
G_{ii}[u] d {\bf x}^2. \eea
Here, $u = u(\rho)$ refers to a functional relation between the size
moduli $\rho$ and the holographic coordinate $u$ to be determined.
By translational symmetry along $\mathbb{R}^3$ and along
$\mathbb{S}^1$, the metric components are functions of $\rho$ only.
Analytic computations of these metric component were not available,
so we extracted them from numerical computation on MATHEMATICA.

We propose to fix function $u(\rho)$ by comparing the information
geometry with the Euclidean anti-de Sitter Schwarzschild black hole.
In terms of Poincar\'e coordinates, the metric is given by
\bea \d s^2 = {1 \over u^2} \left( 1 - {u^4 \over u_0^4} \right) \d
t^2 +  \left( 1 - {u^4 \over u_0^4} \right)^{-1} {\d u^2 \over u^2}
+ {1 \over u^2} \d {\bf x}^2. \label{ssbh2} \eea
The background has a topology of $\mathbb{R}_2 \times \mathbb{R}_3$,
the latter subspace referring to the horizon. In Euclidean
signature, the geometry is complete for $u \ge u_0$, and the
regularity of the $\mathbb{R}_2$ part at $u = u_0$ relates the
temperature and the mass of the black hole. In this Poincar\'e
parametrization, the coordinate $u$ is the radial coordinate and
ranges over $u_0 \le u < \infty$ for Euclidean signature (compared
to the range $ 0 \le u < \infty$ for Lorentzian signature). At
strong `t Hooft coupling regime, leading corrections were computed
in \cite{klebanov}, and indicated that the horizon and the
singularity at $u=0$ persist.

$\bullet$ \underline{$G_{ij}$}: \hfill\break We begin with the
metric component $G_{ij}$, since, according to Eq.(\ref{ssbh2}), we
expect that this metric component is not deformed by the
Schwarzschild harmonic function, viz. takes the same functional form
$1/\rho^2$ as the Euclidean anti-de Sitter space ${\bf H}_5$. In the
previous section, we noted from asymptotic behavior that the caloron
behaves the same as zero temperature Yang-Mills instanton for small
size $\rho \sim 0$, but is gradually deformed for larger size. We
anticipated that, when measured in terms of {\sl zero temperature
instanton size variable} $\rho_T$ in (\ref{caloronrelation}), the
instanton size is bounded above at a finite size set by the
Yang-Mills temperature $T$. Certainly, this is not the behavior one
finds from the $L^2$-metric and ADHM calculus. So, we compared the
relation (\ref{caloronrelation}) with the information metric
component $G_{ij}$.
\begin{figure}
\epsfysize=8cm \centerline{\epsfbox{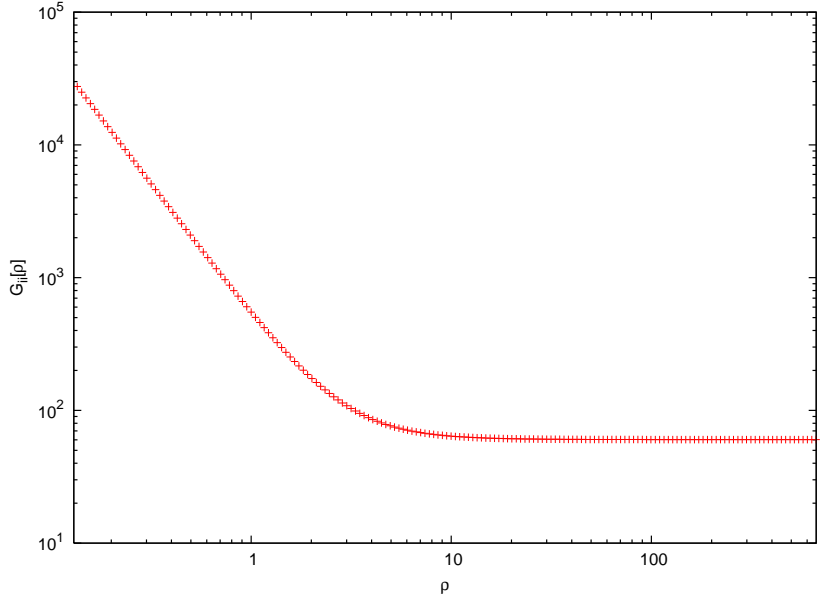}} \caption{\sl
Log-log plot of the information metric $G_{ij}$ as a function of
$\rho$. The numerical data points are cross-marked. The Yang-Mills
gas temperature is set to $T = 1$. Across $\rho = 1/T$, the metric
component $G_{ij}$ behaves very differently.} \label{fig:gii}
\end{figure}
Our computation yields
\bea G_{ij}[\rho] =  506\, \left({1 \over \rho^2} + 0.119 \right)
\delta_{ij}~. \label{gij} \eea
Remarkably, the result displays the same feature as
(\ref{caloronrelation}): $G_{ij}$ approaches to a {\sl constant}
value as $\rho \rightarrow \infty$! Thus, we propose to set the
functional relation $u(\rho)$ as
\bea {1 \over u^2} := \left( {1 \over \rho^2} + {1 \over u_o^2}
\right) \qquad \mbox{where} \qquad {1 \over u_o^2} \sim 0.119 \,
T^2, \label{15} \eea
and identify $u$, not $\rho$, with the radial variable in the metric
(\ref{ssbh2}). Stated differently, (\ref{15}) defines the map
between Yang-Mills theory scale (instanton size) and the emergent
geometry scale (holographic distance). We thus have
\bea G_{ij}[u] = 506 {\delta_{ij} \over u^2}. \eea
Since the radial variable $u$ extends only up to $u_o$, we identify
$u_o \sim 2.9/T$ as the "horizon". Notice that this agrees well with
$2 \sqrt{3}/T$ in (\ref{range}). Accordingly, we interpret the
Yang-Mills gas temperature $T$ as the Hawking temperature.

$\bullet$ \underline{$G_{uu}$}: \hfill\break
Next, consider the metric component $G_{uu}[u]$. From the anti-de
Sitter Schwarzschild metric, we anticipate that $G_{uu}$ diverges at
the "horizon" $u_0$. From Yang-Mills theory side, we will first
compute the metric component $G_{\rho\rho}$ and then change the
variable according to (\ref{15}), convert it to $G_{uu}[u] =
G_{\rho\rho}(\rho) \left({\d \rho \over \d u}\right)^2$.
%
\begin{figure}
\epsfysize=8cm \centerline{\epsfbox{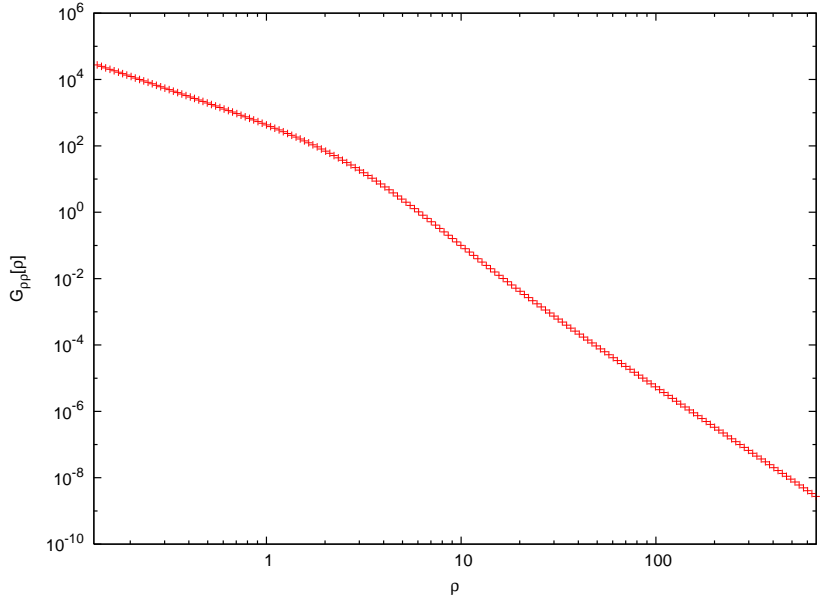}} \caption{\sl
Log-log plot of the information metric $G_{\rho\rho}$. At $\rho \ll
1/T$, it scales as $1/\rho^2$. At $\rho \gg 1/T$, it scales as
$1/\rho^4$. } \label{fig:grr}
\end{figure}
Numerically, we obtained that $G_{\rho\rho}$ interpolates between
$\rho^{-2}$ and $\rho^{-4}$ behaviors as $\rho$ ranges from $0$ to
$\infty$:
\bea G_{\rho\rho}(\rho) =  502 \left( {1 \over \rho^2 + 0.91 \,
\rho^4} \right)~. \eea
The result is plotted in Fig.2.

Changing the variables to $u$, we find that
\bea G_{uu}[u] = 502 {1\over u^2} \Big( 1 - {u^4 \over u_0^4}
\Big)^{-1} R(u) \qquad \mbox{where} \qquad R(u) \sim {1 + u^2/u_0^2
\over 1 +6.70 \, u^2/u_0^2}. \label{grr}\eea
Remarkably, the metric has a simple pole at $u = u_0$, which is
precisely the location of the "horizon".
Notice that the location of the pole originates from the change of
the variables, explaining why it is invariably the same as the
location of the horizon. Compared to the metric (\ref{ssbh2}), we
thus see that the metric component is now modified by the function
$R(u)$. Since these two results are totally opposite regime of the
coupling parameters, we conclude that $R(u)$ summarizes strong
worldsheet and string coupling effects.

$\bullet$ \underline{$G_{tt}$}: \hfill\break
Lastly, we extract $G_{tt}$ component of the information metric.
From the anti-de Sitter Schwarzschild black hole metric, we expect
this component exhibits a single zero at the horizon. From numerical
computation, we have found that the best fit is given by
\bea G_{tt}(\rho) = 507 \left( {1 \over \rho^2 + 0.037
\rho^4}\right)~.\label{gtt} \eea
%
\begin{figure}
\epsfysize=8cm \centerline{\epsfbox{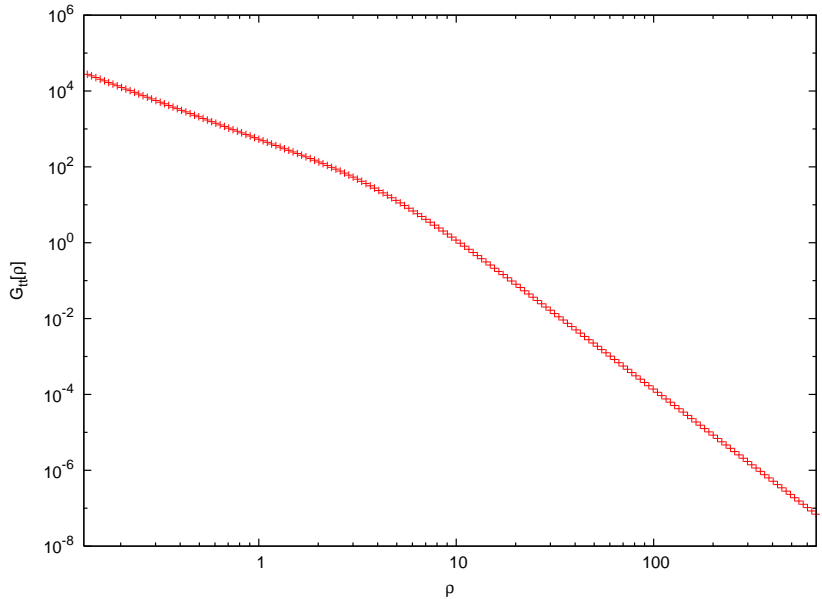}} \caption{\sl
Log-log plot of the information metric $G_{tt}$. At $\rho \ll 1/T$,
it scales as $1/\rho^2$. At $\rho \gg 1/T$, it scales as $1/\rho^4$.
It behaves similar to $G_{\rho\rho}$, but the numerical coefficients
for each slope are different.} \label{}
\end{figure}
Again, changing the $\rho$ variable into $u$ via (\ref{15}), we get
\bea G_{tt}[u] \sim 507 {1 \over u^2} \left(1 - {u^4 \over
u_o^4}\right) T(u) \qquad \mbox{where} \qquad T(u) = {1 -
{u^2/u_o^2} \over 1 - 0.69\,{u^2 / u_o^2}}\left( 1 + {u^2 \over
u_0^2}\right)^{-1}. \eea
We see that the metric vanishes at the "horizon", $u \rightarrow
u_0$! The metric actually exhibits {\sl double} zero at the horizon.
It would be interesting to see if the double zero resolves into a
single zero once numerical accuracy and fitting function with higher
precision are performed, but we relegate the study for future work.
It also exhibits a simple pole, but it is hidden inside the horizon.

Summarizing, the Euclidean metric of the instanton moduli space
takes the form
\bea \d s^2 \sim 500 \left[ {1 \over u^2} \left( 1 - {u^4 \over
u_0^4} \right) T(u) \d t^2 + \left(1 - {u^4 \over u_0^4}
\right)^{-1} R(u) {\d u^2 \over u^2} + {1 \over u^2} \d {\bf x}^2
\right], \label{final} \eea
where $T(u), R(u)$ given in (\ref{gtt}) and (\ref{grr}) represent
effects due to string worldsheet and quantum corrections.

Lorentzian black hole geometry is obtainable by Wick rotating the
Euclidean metric (\ref{final}). The Wick rotation is a well-defined
notion in gauge theory, and this must be hold as well for the
instanton moduli space. It then follows that the Lorentzian black
hole clearly exhibits $u=\infty$ singularity as in the case for the
Schwarzschild black hole. We believe these features bear important
implications to exploring inside the black hole via Maldacena's
AdS/CFT \cite{shenker}. In our numerical computation, there appears
other poles and zeros in the metric components, all located at
finite $u$ and inside the horizon. However, further precision study
is imperative before drawing physical significance of them. Any
extra structures present to the moduli space geometry bear
significant implications to our understanding of string theory since
they represent effects arising from unsuppressed stringy and quantum
fluctuations. Another interesting direction is to reconstruct the
minimal surface of Wilson loop \cite{wl1}-\cite{wl23} directly from
weakly coupled gauge theory. We leave them for future study.

\section*{Acknowledgement}
We thank David J. Gross, Juan Maldacena and Steve H. Shenker for
helpful discussions, and Michael Murray and Nigel Hitchin for
correspondences. This work was supported by KRF BK-21 Physics
Division, KRF Star Faculty Grant, KRF Leading Scientist Grant, and
KOSEF SRC Program Center for Quantum Space-Time (R11-2005-021).


\end{document}